# From Thesis to Transition: An INSIGHT-Inspired Approach to Co-Designing Industry 5.0 Competency Pathways for Early-Stage Researchers

Georg Macher*[2], Omar Veledar[2], Suad Krilašević[10], Raluca Coscodaru[8], Luca Barbera[11], Eleni Matinopoulou[7], Andrea Louzan Perez[3], Sisi Pörröinen-Kaunis[4],Crinela Silvia Dragan[8], Gülsah Solmaz[6], Chiara Ferrarini[5], Franz Valko-Tassu[4], Maryia Petkova[9], Alba Dominguez Pedreira[3], Matteo Falsetta[1]

* Corresponding author, e-mail georg.macher@tugraz.at
[1] EIT Digital, BE
[2] Graz University of Technology, AT
[3] Centro Tecnológico de Automoción de Galicia, ES
[4] Independent researcher, FI
[5] RE:Lab, IT
[6] Fraunhofer Ges. z. Förderung der Angewandten Forschung, DE
[7] Erevnitiko Panepistimiako Institouto Systimation Epikoinonion Kai Ypologiston, EL
[8] Unitatea Executiva Pentru Finantarea Invatamantului Superior A Cercetarii Dezvoltarii Si Inovarii, RO
[9] Regional Agency for Entrepreneurship and Innovations Varna, BG
[10] Asocijacija za napredak nauke I tehnologije, BA
[11] University Industry Innovation Network, NL

**Abstract:** Europe faces a critical "translation gap" where doctoral excellence in academia often fails to convert into industrial impact. While Industry 5.0 demands a blend of technical depth, sustainability, and human-centric design, traditional higher academic education remains siloed. This paper presents an approach from the Horizon Europe INSIGHT initiative to co-design modular competency pathways for early-stage researchers. Using a multi-methodological analysis framework, including expert interviews and co-design workshops, we propose a two-layer competency architecture. This layers foundational translational skills (communication, project management) with Industry 5.0 literacies (data governance, value creation). Rather than proposing fixed training tracks, the paper outlines emerging pathway directions and the design principles behind them: modularity, practical relevance, mentoring-rich support, and cross-sector applicability. Its contribution is a scalable way of researcher development towards useful real-world application.

**Keywords:** Industry 5.0; Early-stage researchers; Competency pathways; Cross-sector collaboration; Transferable skills; Innovation management.

## 1 Introduction

Europe does not lack talented researchers. Doctoral students are vital in network configurations, acting as primary vessels for knowledge transfer between universities and companies. Their research experience is increasingly conditioned by the degree of institutionalisation of collaborative arrangements and the prior experience of the involved parties (Thune, 2009). What Europe lacks is enough researchers who can move effectively between academic environments and application-oriented innovation settings. In many disciplines, doctoral and early research training still produce strong technical depth, methodological rigour, and scientific originality, yet these strengths do not automatically translate into cross-sector readiness. The problem is not that researchers are underqualified. It is that the environments in which they are trained and the environments in which they are later expected to create value often reward different things. Academic contexts still tend to privilege novelty, publications, and disciplinary excellence, while industry and broader innovation settings often prioritise implementation, reliability, speed, collaboration, and usable solutions. As a result, many early-stage researchers enter the labour market with impressive scientific capability but weaker preparation for navigating non-academic expectations, translating research into practical value, and building careers across sectors. While doctoral expansion aligns with the rise of the knowledge economy, employment outcomes are highly differentiated by subject area and doctoral institution. Statistics indicate that while two-thirds of UK PhD graduates enter non-academic employment, securing specific research roles is significantly more common for science graduates (Hancock, 2023).

This tension becomes more dominant in the context of Industry 5.0. Here, technical knowledge remains important, but it is no longer sufficient on its own. Researchers increasingly need to operate at the intersection of digital technologies, sustainability, and human-centred design, while also engaging with stakeholders beyond their own disciplinary communities. They may be expected to understand not only methods and results, but also collaboration across organisational cultures, technology acceptance, implementation constraints, business relevance, and the broader societal implications of innovation. Early exploratory inputs gathered in the INSIGHT project[1] reinforce this view. Across questionnaires, interviews, workshops, and project discussions, the recurring signals point not simply to missing technical competences, but to a broader need for translational ability, interdisciplinary communication, project and stakeholder coordination, mentoring support, and stronger understanding of how research becomes something organisations can use, adopt, and value.

This suggests that the challenge is better understood not only as a skills gap, but as a **translation gap** between academic logic and application-oriented environments. That distinction matters. If the issue is framed too narrowly, the response tends to become equally narrow: a few add-on soft-skills sessions, some generic employability advice, or isolated training modules detached from the realities researchers actually face. The evidence considered in this paper points in a different direction. What seems to be needed is a more structured and modular approach that combines transferable competences with industry-facing and Industry 5.0-relevant implementation capabilities, while also taking seriously the role of mentoring, practical exposure, and learning formats that connect training to real contexts. In other words, the problem is not solved by telling researchers to become “more employable”. It requires a clearer architecture for how academic capability can be translated into cross-sector impact.

[1] https://insighthorizon.eu/

Research suggests that PhD graduates' transition to industry is increasingly mediated by autonomously built personal networks that match scientific expertise with market demand. These university-industry connections play a vital role in reducing search costs and uncertainty during the job search process (Germain-Alamartine et al., 2021). Against this background, this paper presents an **INSIGHT-inspired approach** for co-designing modular competency pathways that support early-stage researchers in moving from academic capability towards cross-sector, Industry 5.0-relevant impact. Rather than proposing fixed training tracks, the paper introduces a two-layer competency architecture, shaped through multiple exploratory evidence streams and uses it to outline emerging pathway directions. The first layer covers transferable and translational competences such as communication, collaboration, project management, and career navigation. The second covers industry-facing and Industry 5.0 implementation competences such as human-centred design, digital and data literacy, sustainability, exploitation and value creation, and increasingly, trustworthy and governance-aware technology use. The contribution of the paper lies in offering a scalable way to think about researcher development beyond isolated skills training, with relevance not only for the INSIGHT project, but also for innovation managers, university-industry intermediaries, and organisations seeking to better align doctoral and early-career talent with the realities of contemporary innovation ecosystems.

## 2 Background: Why a Pathway Approach is needed

The gap between doctoral training and labour market needs is not a new concern, but in practice it remains stubbornly unresolved. Research training continues to produce strong disciplinary expertise, analytical depth, and methodological rigour, yet these strengths do not automatically prepare researchers for environments in which value is judged through implementation, coordination, business relevance, or cross-functional contribution. The tension becomes especially visible in settings where researchers are expected to move between university and industry, or where industrially relevant work must still satisfy academic expectations around publication, formal progression, and disciplinary legitimacy. Several of the inputs considered in this paper describe exactly this problem: researchers caught between systems that both value their work, but value it differently. University-industry collaborations for R&D&I face diverse barriers related to cultural conflicts, intellectual property issues, and misaligned research timelines. These obstacles can be mitigated by fostering relational social capital and gradually increasing the complexity of collaborative projects over time (Rossoni et al., 2024).
Seen from that perspective, the issue is not simply that researchers “lack practical skills”. That explanation is too narrow and, in some cases, misleading. The material analysed here points to a broader mismatch between academic preparation and cross-sector contribution. What is often missing is not intelligence or technical competence, but the ability to communicate across boundaries, understand non-academic expectations, translate research into forms that others can use, and operate within organisational contexts shaped by deadlines, stakeholder interests, implementation constraints, and different definitions of success. This is why the problem is better described as a transition challenge, or more precisely, as a translation challenge between research logic and application logic.
Industry 5.0 makes this challenge sharper rather than softer. It acts as a policy framework intended to govern and regulate the trajectory of Industry 4.0 toward societal and environmental goals. Realising this agenda requires leveraging digital manufacturing functions (e.g., business risk management and smart circular products) in contextual sequences that adapt to turbulent business uncertainties (Ghobakhloo et al., 2024). It raises expectations in several directions at once. Researchers are no longer asked only to understand technologies in isolation. They are increasingly expected to work at the intersection of digital systems, data-rich environments, sustainability demands, human-centred design, and organisational adoption realities. In other words, readiness for contribution now includes technical literacy, but also the ability to understand people, systems, use contexts, and implementation consequences. The **workshop and questionnaire outcomes of INSIGHT** both support this broader reading, highlighting themes such as human-centred technology, digital and data fluency, sustainability, technology acceptance, and the need to work across disciplinary and institutional boundaries.
The more industry-near inputs refine this point further. These **facilitation activity outcomes** suggest that contemporary readiness is not only about knowing what new technologies can do, but also about judging whether they are useful, reliable, secure, and appropriate in organisational settings. Concerns around practical adoption, verification of AI-supported outputs, privacy, security, data integrity, and infrastructure choices show that implementation capability now includes governance-aware judgement as well as technical understanding. This matters because it changes the profile of the "future-ready researcher". The challenge is no longer only to educate specialists who can create knowledge, but also to support people who can move that knowledge into environments where trust, risk, responsibility, and operational reality shape what becomes possible.
If that diagnosis is correct, then fragmented responses are unlikely to be enough. A few isolated workshops, generic employability sessions, or add-on modules may help at the margins, but they do not offer a strong enough response to a problem that spans competencies, contexts, and learning conditions. **What is needed instead is a more deliberate architecture for development**, one that connects transferable capabilities, industry-facing literacies, learning formats, mentoring, and practical exposure. This is the reason for adopting a pathway perspective in this paper. A pathway approach makes it possible to move beyond undifferentiated lists of desirable skills and towards a more structured model of how researchers can be supported in progressing from academic capability towards wider forms of contribution.
In that sense, the background to this paper is not only a problem of training content. It is also a problem of design. The central question is not simply what early-stage researchers should know, but how their development should be structured so that scientific depth can connect more effectively with cross-sector collaboration, innovation processes, and Industry 5.0-oriented implementation. The next section responds to that question by presenting the INSIGHT-inspired approach used to begin organising this design space.

## 3 The INSIGHT-inspired Approach

The approach presented in this paper is designed to support early-stage researchers in moving from academic capability towards cross-sector, Industry 5.0-relevant impact. Its underlying premise is simple: useful competency pathways should not be defined in advance as a fixed curriculum. They should emerge from a structured process that combines different kinds of insight about researcher needs, industry expectations, practical constraints, and future-oriented capability demands. The aim is therefore not to prescribe a universal training model, but to build a scalable approach for identifying what researchers need in order to work more effectively across academic and application-oriented settings.

To do this, the approach combines several complementary evidence streams. These are not treated as equivalent datasets, nor are they intended to produce a single statistically unified picture. Instead, they are used to surface recurring patterns, tensions, and priorities that can guide the design of modular competency pathways. In this sense, the approach is better understood as **evidence-informed co-design** than as a conventional needs assessment alone.

The **first** evidence stream is **structured scoping**. This was used to organise the competency space into a set of broad but actionable domains. These include transferable competences related to impact, research management, and collaboration, as well as more specific themes linked to human-centred technology, digital and data-related work, sustainability, and learning preferences. The value of this scoping step is not that it produces final answers, but that it makes the problem more visible and comparable across respondents and contexts.

The **second** evidence stream is **expert deepening**. Interviews add practical and interpretive depth by revealing issues that structured instruments tend to miss. These include the importance of strategic thinking, leadership, relevance to daily work, understanding one's value in non-academic settings, and the need to make training directly applicable to real organisational contexts. Such insights help move the discussion beyond abstract skill categories and towards the lived realities of transition between research and practice.

The **third** evidence stream is **interactive discussion**. Workshop-based conversations make it possible to test assumptions in a more open way and to capture how participants themselves frame the gap between academia and industry. University-industry collaborative education can function as a "trading zone" where theoretical insights from academia and experiential know-how from industry are efficiently exchanged. This exchange is driven by the heterogeneity between participants, requiring program facilitators to act as mediators who establish a shared language (Nakagawa et al., 2017). Across such discussions, several themes recur: tension between publication logic and industrial deliverables, the centrality of communication and interdisciplinary cooperation, the value of mixed-background role models, and the need for researchers to learn how to move between different expectations, languages, and forms of value. These discussions are especially useful because they expose not only desired skills, but also structural frictions and practical misunderstandings.

A **fourth**, more auxiliary stream can be described as **market-near validation**. Feedback from learners already embedded in industrial environments helps refine the application-facing side of the approach. In particular, it strengthens the importance of practical technology adoption, trustworthy use of AI, verification of outputs, security and privacy, data integrity, and governance-aware implementation. This does not replace the perspective of early-stage researchers, but it helps ensure that the resulting competency model is not overly academic in tone or overly detached from organisational reality.

Taken together, these evidence streams support a **two-layer competency architecture**. The **first layer covers transferable and translational competences**, such as communication across audiences, interdisciplinary collaboration, project and resource management, self-management, strategic thinking, and career navigation. These are the capabilities that allow researchers to operate across settings rather than only within their own disciplinary niche. The **second layer covers industry-facing and Industry 5.0 implementation competences**, such as human-centred design, technology acceptance, digital and data literacy, sustainability, systems thinking, exploitation and value creation, and increasingly, trustworthy and governance-aware technology use. The practical value of separating these layers is that it avoids collapsing everything into either vague soft skills or narrow technical content. Figure 1 provides an overview of the INSIGHT-inspired approach, linking evidence streams, competency architecture, design principles, and emerging pathway directions.

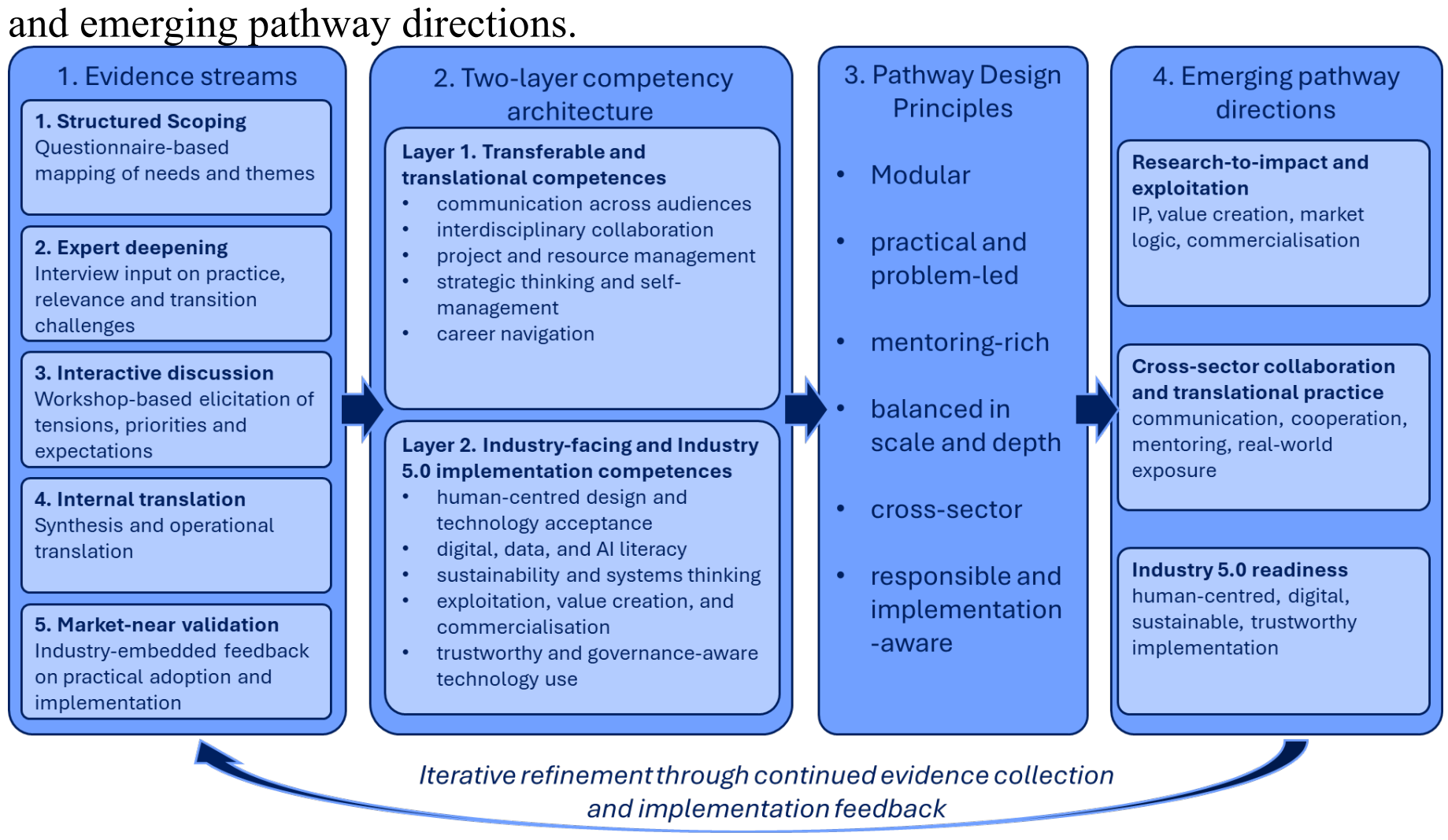

**Figure 1** INSIGHT-inspired approach for co-designing Industry 5.0 competency pathways.

As shown in Figure 1, the value of the approach lies not in any single source of input, but in the way different evidence streams are translated into a structured and adaptable pathway logic. This architecture also suggests a small set of pathway design principles. **First**, pathways should be **modular**, because needs differ by sector, role, and prior experience. **Second**, they should be **practical and problem-led**, since multiple sources point to the importance of real cases and direct relevance to work. Structured innovation training that balances practical knowledge with enjoyable learning experiences can anchor creative confidence and competence in learners over the long term. Practical relevance is enhanced by adapting real-life innovation project cases, which encourages learners to actively construct meaningful interpretations (Reis & Hunt, 2017). **Third**, they should be **mentoring-rich**, because credible guidance from mixed-background professionals appears more useful than formal training alone. While many faculty members have prior industry experience, this does not always translate into high-quality transferable skills instruction without intentional pedagogical strategies. Such experience provides specific "cultural scripts" that emphasise real-world practicality and divergent thinking, which can be effectively shared via faculty development programs (Hora & Lee, 2024). **Fourth**, they should **balance scale and depth**, combining broader-access formats with smaller, more intensive forms of learning where needed. Finally, they should prepare researchers not only to acquire competences, but also to navigate cross-sector contexts and make responsible judgements about implementation in real organisations.

In that sense, the INSIGHT-inspired approach has collected multi-stakeholder opinions about trainings and educational models. It is a way of structuring an evolving design problem by creating a basis for pathway development that is specific enough to guide action, but flexible enough to evolve. The next section builds on this foundation by examining the main themes that emerge across the evidence streams and the pathway directions they begin to suggest.

## 4 Emerging Findings from the Evidence Streams

Taken together, the evidence streams point to a fairly consistent conclusion: the gap between academic training and industry-facing work is not simply a matter of missing practical skills. It is better understood as a broader translation problem. Researchers are often expected to move between environments that differ in language, incentives, measures of success, and ways of defining value. Academic settings tend to reward methodological rigour, originality, and publication output, while application-oriented environments often prioritise implementation, reliability, speed, and usefulness. Several inputs describe researchers finding themselves caught between these expectations, especially in contexts where university and industry logics coexist but are not well aligned. This suggests that any serious **pathway design** must prepare researchers not only to know more, but also to translate, negotiate, and operate between different systems of work.

A **second strong finding** is that **transferable competences are central** rather than secondary. Across the sources, communication, interdisciplinary cooperation, teamwork, project and resource management, strategic thinking, and the ability to work with different stakeholders recur with remarkable consistency. They do not appear as vague "soft skills" added at the margins, but as conditions for making technical expertise useful outside narrow disciplinary settings. This is particularly visible in the workshop responses, where communication across disciplines and cooperation in multidisciplinary contexts are repeatedly highlighted, and in the interview material, where leadership, ownership, and the ability to make meaningful contributions are treated as decisive for performance and credibility. In complex organisational settings, informal leadership is a key practice for maximising the impact of collaborative projects through variability and reciprocity in leading. Impact is maximised when individuals invest in human relations and shared strategic goals, facilitating knowledge sharing and coherence across networks (Leino, et al., 2022). In practical terms, this means that researcher development cannot be reduced to content exposure alone. It must also include the capabilities needed to make knowledge travel across boundaries.

A **third finding** is that **exploitation and value translation** remain weakly embedded areas. Multiple sources point to uncertainty around intellectual property, patents, ownership, publication timing, market awareness, go-to-market logic, pitching, and the general problem of turning research into something organisations or markets can recognise as valuable. This does not mean that all early-stage researchers need to become founders or business developers. It does suggest, however, that many are underprepared to understand how research can be positioned, protected, communicated, and developed beyond publication. The recurring references to business modelling, market analysis, commercialisation, and start-up feedback indicate that this is not a niche concern. It is part of what it now means to prepare researchers for broader innovation ecosystems.

A **fourth finding** is that **Industry 5.0 broadens the meaning of research readiness**. The evidence does not support a narrow interpretation centred only on digital tools or technical updating. Instead, it points towards a combination of human-centred design, technology acceptance, digital and data-related literacy, sustainability, systems thinking, and sensitivity to societal and environmental implications. The recurring emphasis on human interaction with technology, trust, interdisciplinary cooperation, and environmental impact suggests that researchers are increasingly expected to understand how technologies function within larger technical and social systems, not only within a laboratory or disciplinary context. This broadening of readiness is important because it shows why conventional doctoral training, even when technically strong, may still leave important blind spots.

A **fifth finding**, strengthened especially by the more industry-embedded feedback, is that **practical implementation now also requires trustworthy and governance-aware judgement**. Participants do not simply ask for more exposure to AI, machine learning, or large language models. They repeatedly ask how to verify outputs, assess whether AI-supported analysis is reliable, manage prompt security, protect privacy, preserve data integrity, and make choices about infrastructure and compliance. The concurrent enhancement of soft skills and digital competencies is markedly beneficial for student preparedness in labour markets transformed by AI tools. In these environments, student autonomy serves as a vital mediator in converting the utilisation of GenAI tools into meaningful academic and career outcomes (Hrnjic et al., 2025). Hence, the issue is no longer only whether researchers understand advanced technologies, but whether they can judge when those technologies are useful, trustworthy, and appropriate in organisational settings. This gives the industry-facing competence layer a sharper and more contemporary meaning. Challenging hierarchical structures allows technical experts to influence decision-makers and boost technical outcomes within network-based environments. Practices such as taking responsibility and information brokering enable researchers to interweave administrative context with technical knowledge (Leino et al., 2022). It is not only about adoption, but about responsible and critical implementation.

A **sixth finding** is that **effective support is likely to require more than formal training** provision. Across the interview, workshop, and discussion material, there is strong support for mixed-background mentors, hybrid role models, postdoctoral staff with industrial experience, short placements or visits, and learning formats grounded in real cases. The consistent message is that credibility matters. Researchers appear to benefit not only from being told what matters, but from seeing how experienced people navigate the boundary between research and application in practice. This is one reason why mentoring emerges so strongly across the evidence: it offers access to tacit knowledge, situated judgement, and practical translation that standard training sessions may struggle to deliver on their own. Distributed peer mentoring networks (DPMNs) offer virtual environments where ECRs exchange social capital and research ideas across borders and hierarchies. Participation in such networks has been found to boost individual productivity and provide essential emotional and motivational support (Martin et al., 2023).

A **seventh finding** is that **delivery format and treatment style matter strategically**. Participants across different contexts consistently favour interactive, application-based, and case-led formats over generic or purely expository teaching. Online sessions and self-paced materials appear useful for reach and flexibility, while bootcamps, smaller group formats, and discussion-based work seem better suited to deeper engagement and actual production of ideas or outputs. At the same time, the more industry-near feedback shows clear impatience with repetitive coverage, broad technology hype, and technical detail that lacks decision relevance. This suggests that format cannot be separated from pedagogy. A useful pathway does not only choose the right topics. It must also decide what level of depth is needed, for whom, and in which form the content becomes credible and usable.

Taken together, these findings support a view of researcher development that is broader, more applied, and more structurally aware than many conventional training models. The evidence does not point towards a single missing competence that can be fixed with one intervention. It points instead towards a set of interacting needs: translation between contexts, stronger transferable competences, clearer pathways to exploitation and value creation, broader Industry 5.0 implementation literacy, and support mechanisms that combine training with mentoring and practical exposure. On that basis, the next step is not to prescribe a finished curriculum, but to identify the pathway directions that begin to emerge from this combination of signals.

## 5 Emerging Pathway Directions

At this stage, the evidence does not justify presenting fixed training tracks. The pathways are still being shaped, and the value of the **current work lies more in clarifying the direction of travel** than in claiming a finished model. What does emerge with reasonable consistency, however, is a set of pathway directions that can help organise future development. These directions are not intended as rigid categories. They are better understood as modular orientations that can be combined differently depending on context, sector, and participant need.

A **first direction** can be described as **research-to-impact and exploitation**. This direction responds to the recurring signal that many researchers are insufficiently prepared to understand how research creates value outside publication. The evidence points repeatedly to uncertainty around intellectual property, patents, ownership, publication strategy, market logic, pitching, and commercialisation-related thinking. It also shows interest in business modelling, go-to-market strategy, spin-off creation, and understanding how technical work can be translated into something that organisations or markets can recognise as useful. A pathway built around this direction would therefore not train researchers to abandon science for business. Rather, it would help them understand how research findings can be positioned, protected, communicated, and developed in ways that increase their practical and economic relevance.

A **second direction** can be described as **cross-sector collaboration and translational practice**. This pathway responds to the strong evidence that the boundary between academic and non-academic contexts is difficult to navigate not only because of missing content knowledge, but because of weak preparation for different organisational logics, expectations, and working cultures. Communication across disciplines, collaboration with non-experts, handling different incentives, and understanding the distinction between scientific and application-oriented value all appear as recurring themes. The evidence also suggests that this type of readiness is best supported not only through formal training, but through mixed mentoring, multidisciplinary teamwork, short visits, and practical exposure to real problems. A pathway built around this direction would therefore focus on helping researchers move between worlds: to communicate, cooperate, negotiate, and contribute in settings where academic excellence alone is not enough. The "embedded researcher" model illustrates the value of situating researchers within organisations to co-produce timely evidence aligned with organisational priorities. Supporting ECRs in such roles requires focused training in leadership and organisational management to navigate misaligned priorities between workplace cultures (Chukwu et al., 2024).

A **third direction** can be described as **Industry 5.0 readiness**. Industry 5.0 distinguishes itself from Industry 4.0 by prioritising three core pillars: human centricity, sustainability, and resilience. In this paradigm, human-machine interaction shifts so that the human being controls and uses technology as a resource adaptable to individual worker needs (Alves et al., 2023). This third pathway reflects the broadening of research capability into a wider implementation context shaped by digitalisation, sustainability, and human-centred design. The evidence points towards a need for competences related to human-technology interaction, technology acceptance, data and digital literacy, systems thinking, sustainability, and more recently, trustworthy and governance-aware use of advanced technologies such as AI. What is notable here is that the relevant content is not purely technical. It includes the ability to judge where technologies fit, how they should be introduced responsibly, and how organisational constraints such as security, privacy, data integrity, and compliance influence adoption. A pathway built around this direction would therefore prepare researchers not only to understand emerging technologies, but also to place them in realistic organisational, social, and environmental contexts.

These three directions are analytically distinct, but they are not meant to operate in isolation. In practice, the strongest pathway designs are likely to combine them. A researcher interested in applied AI, for example, may need elements of Industry 5.0 readiness, but also support in collaboration across sectors and some understanding of value creation or intellectual property. Another researcher may need primarily translational and mentoring support, but also enough sustainability or human-centred literacy to contribute credibly in an Industry 5.0 environment. This is why modularity matters. The evidence does not support a one-size-fits-all curriculum, but it does support a structured architecture in which pathway components can be assembled around common directions.

What is important at this stage is therefore not the final naming or packaging of the pathways, but the fact that the evidence is already helping to reduce complexity. Instead of treating researcher development as a long undifferentiated list of desirable skills, the emerging directions create a more manageable design space. They suggest that pathway development can proceed by asking a clearer set of questions: How much emphasis should be placed on exploitation and market-facing capability? How much on collaboration across contexts? And how much on readiness for human-centred, digital, sustainable, and trustworthy implementation? The next phase of development will refine these balances further. For now, the main contribution is to show that the pathway logic is beginning to take shape in a way that is structured enough to guide action, yet still flexible enough to evolve.

The next section discusses what these emerging directions imply more broadly for researcher development, innovation management, and the challenge of preparing talent for Industry 5.0-oriented ecosystems.

## 6 Discussion

The evidence considered in this paper suggests that the widely used idea of a "skills mismatch" is both useful and incomplete. It is useful because there are indeed identifiable areas in which early-stage researchers may be underprepared for work outside narrow academic settings. Communication across domains, project coordination, value creation, intellectual property, and practical technology adoption all emerge as areas requiring stronger support. At the same time, the phrase becomes misleading if it implies that the problem can be solved simply by adding more topics to a training catalogue. What the evidence shows more clearly is **a broader translation gap**. Employability agency (facilitated by agentic dispositions like persistence and self-determination) is a significant factor in successful career transitions for PhD holders. However, the intellectual knowledge acquired during a PhD is often insufficient on its own, as its application is frequently determined by the graduate's social and cultural capital (Pham, 2025). Researchers are often expected to move between environments that define success differently, speak different organisational languages, and attach value to different outputs. Discrepancies in skill demand between industries are often shaped by deep-seated cultural norms and guiding values rather than just a simple lack of skills. Effective employability training must therefore move beyond "umbrella terms" to incorporate a sociological perspective that acknowledges how industry value systems dictate skill implementations (Chen et al., 2026). In that sense, the real challenge is not only capability acquisition, but the ability to recognise, interpret, and navigate contrasting systems of work.

This shift in emphasis matters for innovation specialists. Innovation systems do not depend only on technical novelty. They depend on people who can connect research and application, explain complex ideas across boundaries, recognise where knowledge can become useful, and work credibly with actors who do not share the same assumptions about evidence, speed, risk, or value. From that perspective, **the contribution of this paper** is less about proposing another employability framework and more **about suggesting a way to structure researcher development around these bridging functions**. For innovation managers, HR professionals, university-industry intermediaries, and technology transfer actors, the relevance is practical: if talent development is treated only as specialist upskilling, a large part of innovation potential remains trapped in translation failure. A pathway-based approach offers a more actionable alternative because it links competence areas, mentoring, learning formats, and practical exposure instead of treating them as separate concerns.

A further implication of the findings is that industry-facing readiness increasingly includes responsible and governance-aware implementation. This becomes especially visible when the evidence is broadened beyond researcher-centred material to include more market-near perspectives. The MBA feedback, in particular, reinforces that practical organisational value is not created by technology enthusiasm alone. Participants consistently ask how to verify AI-supported outputs, manage privacy and security, preserve data integrity, and make realistic choices about infrastructure, compliance, and control. This suggests that the **second competency** layer proposed in this paper should not be understood narrowly as "technical literacy for industry", but **more broadly as implementation literacy**: the ability to assess whether a technology is useful, trustworthy, and appropriate in a given organisational context. For Industry 5.0, this is especially important because advanced technologies increasingly enter environments where human, regulatory, and ethical considerations are inseparable from technical function.

The paper also points towards a useful way of thinking about generalisability. Although the present work is grounded in an INSIGHT-inspired process, its value does not depend on replication of the exact project context. The most transferable element is not a fixed set of modules, but the underlying logic: combine multiple exploratory evidence streams, distinguish between transferable and implementation-facing competences, and use that architecture to develop modular pathways rather than a one-size-fits-all curriculum. This logic should be adaptable across sectors, institutional settings, and national contexts. In some environments, the emphasis may fall more strongly on exploitation and entrepreneurship; in others, on interdisciplinary collaboration, sustainability, or trustworthy digital implementation. The strength of the approach lies precisely in its ability to accommodate such variation without losing structural clarity.

At the same time, the present paper should be read in light of clear limitations. The evidence base is exploratory and heterogeneous by design. It includes a draft scoping instrument, a still-limited number of interviews, workshop-based discussion material, internal project reflection, and an auxiliary market-near validation stream. These sources are useful in combination, but they do not provide the kind of standardised empirical base that would justify strong claims of validation or general statistical representativeness. In addition, the **pathway directions presented here remain provisional and are still being refined. This is not a finished model of researcher development, nor is it intended to be one at this stage.** Its contribution is earlier and more strategic: to clarify how pathway design can be approached in a way that is structured enough to guide action, yet flexible enough to learn from continued evidence and implementation.

Taken together, these points suggest that the main value of the work is not in having already solved the transition problem, but in making it more tractable. By moving from a vague discussion of employability or skills gaps toward a layered, modular, and co-designed view of researcher development, the paper offers a more usable way to think about how early-stage researchers can be supported in crossing the boundary between academic excellence and broader innovation impact. The final section draws these strands together and summarises the contribution of the paper.

## 7 Conclusion

The real bottleneck is not at the level of talent production. Universities are already producing highly capable researchers. The **bottleneck appears** later, at the point **where that capability is expected to travel**: into organisations, across sectors, through collaboration, and towards outcomes that matter beyond publication. What this paper suggests is that the distance between research and impact is sustained less by a lack of brilliance than by weak preparation for movement between contexts.

That is why the usual response feels too small. A little more career advice, a workshop on communication, or a generic employability add-on may be useful, but none of them is equal to the problem. What is needed is a more **deliberate structure for transition**: one that recognises that researchers must learn not only to generate knowledge, but also to position it, translate it, defend it, adapt it, and connect it to settings where value is judged differently. The INSIGHT-inspired approach proposed here is an attempt to give that transition a clearer architecture through modular pathway design grounded in transferable, translational, and implementation-facing competences.

Its importance lies less in having produced a finished model than in making the design space more workable. Instead of treating researcher development as a long and unfocused list of desirable traits, the **paper narrows attention to three emerging directions**: (a) research-to-impact and exploitation, (b) cross-sector collaboration and translational practice, and (c) Industry 5.0 readiness. This turns an abstract ambition into something that can be shaped, tested, and improved. For innovation-oriented organisations, this is not a peripheral educational question. It is a question of whether valuable knowledge stays trapped in its original setting or becomes capable of doing work elsewhere.

The implication is straightforward. If we want better innovation outcomes, we should stop treating researcher transition as something that will somehow sort itself out. It needs to be designed with the same seriousness that we apply to technology, partnerships, and strategy. The next task is not to describe the gap more elegantly. It is to build pathways that are credible enough to follow, practical enough to use, and bold enough to change where research can go.

## Acknowledgement

This research has received funding from the European Union under Grant Agreement No 101217020. Views and opinions expressed are those of the author(s) only and do not necessarily reflect those of the European Union or the granting authority.

This work was conducted with the assistance of AI tools, which served to support various stages of the research, such as information synthesis and data interpretation.

Additionally, AI writing assistance was used as a tool for author development, providing feedback on writing style and structure.

The authors believe that this approach has enhanced both the quality of the paper and their own research communication skills, while the authors maintain full responsibility for the content and accuracy of the information presented.